\documentclass[fleqn,usenatbib]{mnras}

\usepackage[T1]{fontenc}
\usepackage{ae,aecompl}

\usepackage{graphicx}	
\usepackage{amsmath}	
\usepackage{amssymb}	
\usepackage{float}
\usepackage{fancyhdr}
\usepackage{hyperref}
\usepackage{longtable}
\usepackage{graphicx}
\usepackage[usenames,dvipsnames]{xcolor}

\usepackage{soul}

\usepackage{epstopdf}

\newcommand{\Teff}{\mbox{$T_{\mathrm{eff}}$}}

\newcommand{\Msun}{\mbox{$\mathrm{M}_{\odot}$}}

\newcounter{tref}

\title[Discovery of the first resolved triple white dwarf]{Discovery
  of the first resolved triple white dwarf}

\author[M. Perpiny\`a-Vall\`es et al.]
{M. Perpiny\`a-Vall\`es$^{1,2}$, A. Rebassa-Mansergas$^{1,3}$, B. T. G\"ansicke$^{2}$, S. Toonen$^{4}$,\newauthor 
J.~J.~Hermes$^{5}$\thanks{Hubble Fellow}, N. P. Gentile Fusillo$^{2}$, P.-E. Tremblay$^{2}$ 
\\
$^{1}$Departament de F\'{\i}sica, Universitat Polit\`{e}cnica de Catalunya, c/Esteve Terrades 5, 08860 Castelldefels, Spain\\
$^{2}$Department of Physics, University of Warwick, Coventry CV4 7AL, UK\\
$^{3}$Institut d'Estudis Espacials de Catalunya, Ed. Nexus-201, c/Gran Capit\`a 2-4, 08034 Barcelona, Spain\\
$^{4}$Anton Pannekoek Institute for Astronomy, University of Amsterdam, 1090 GE Amsterdam, The Netherlands\\
$^{5}$Department of Physics and Astronomy, University of North Carolina, Chapel Hill, NC 27599, USA
}

\date{Accepted XXX. Received YYY; in original form ZZZ}

\pubyear{2018}

\begin{document}
\label{firstpage}
\pagerange{\pageref{firstpage}--\pageref{lastpage}}
\maketitle

\begin{abstract}
We report the discovery of J1953$-$1019, the first resolved triple
white dwarf system. The triplet consists of an inner white dwarf
binary and a wider companion. Using \emph{Gaia} DR2 photometry and
astrometry combined with our follow-up spectroscopy, we derive
effective temperatures, surface gravities, masses and cooling ages of
the three components. All three white dwarfs have pure-hydrogen (DA)
atmospheres, masses of $0.60-0.63\,\Msun$ and cooling ages of
$40-290$\,Myr. We adopt eight initial-to-final mass relations to
estimate the main sequence progenitor masses (which we find to be
similar for the three components, 1.6-2.6\,\Msun) and lifetimes. The
differences between the derived cooling times and main sequence
lifetimes agree for most of the adopted initial-to-final mass
relations, hence the three white dwarfs in J1953$-$1019 are consistent
with coeval evolution. Furthermore, we calculate the projected orbital
separations of the inner white dwarf binary ($303.25\pm0.01$\,au) and
of the centre of mass of the inner binary and the outer companion
($6\,398.97\pm0.09$\,au). From these values, and taking into account a
wide range of possible configurations for the triplet to be currently
dynamically stable, we analyse the future evolution of the system. We
find that a collision between the two inner white dwarfs due to
Lidov-Kozai oscillations is unlikely, though if it occurs it could
result in a sub-Chandrasekhar Type Ia supernova explosion.
\end{abstract}




\section{Introduction}
\label{sec:intro}

White dwarfs are the remnants of main sequence stars of masses up to
$\simeq8-10$\,\Msun\, \citep{Althaus2010a}. Once all the nuclear
burning phases that drive the evolution of these low-to-intermediate
mass main sequence stars come to an end, they lose their outer layers
by stellar winds or stellar interactions leaving behind their remnant
hot, compact cores after the planetary nebula phase. These inert hot
stellar remnants, supported by the pressure of the degenerate
electrons in their interiors, then enter the white dwarf cooling
sequence where their physical characteristics can be studied in
detail. The gravothermal cooling process of white dwarfs is moderately
well understood \citep{althaus+benvenuto98-1, fontaineetal01-1,
  renedoetal10-1, Camisassa2016}, which turns white dwarfs into
``cosmochronometers'' that can be used to, e.g., derive the age of the
Galactic disk \citep{wingetetal87-1, GB88b, Gianninas2015, Kilic2017},
the ages of both open and globular clusters \citep{Calamida2008,
  GBerro2010, Torres2015}, or to study the kinematic properties of the
Galaxy through the age-velocity dispersion relation
\citep{Anguiano2017}.

White dwarfs are not only found as isolated stars but also in binary
and multiple stellar systems \citep{toonenetal17-1}. In fact, the
first two white dwarfs discovered are members of a triple \citep[40
  Eri B;][]{Herschel1785, Bond2017a} and a binary system \citep[Sirius
  B;][]{Bessel1844, Bond2017b}, respectively. White dwarf binaries
have played a key role in constraining a wide variety of open problems
in astrophysics, such as the evolution of close compact binaries
\citep{zorotovicetal10-1, Rebassa2012, Camacho2014, Zorotovic2014},
the age-metallicity relation of the Galactic disk \citep{Zhao2011,
  Rebassa2016}, the age-rotation-activity relation of low-mas main
sequence stars \citep{Rebassa2013, Skinner2017}, the pairing function
of main sequence binaries \citep{Ferrario2012, Cojocaru2017}, and the
initial-to-final mass relation \citep{Catalan2008a, girvenetal10-1,
  Zhao2012, Baxter2014, andrewsetal15-1}. Last, but not least white
dwarf binaries are the progenitors of type Ia supernovae
\citep[SN\,Ia;][]{WangHan12, Maoz2014, Soker2018}, which are key
distance beacons for cosmological studies.

White dwarfs in multiple (at least three components) systems are
relatively rare compared to those in binaries. In the solar
neighbourhood (within 20\,pc), only nine such systems have been so far
identified \citep[see][and references therein]{toonenetal17-1}, and
all of them contain at least one non-degenerate component. The only
systems currently known that contain three compact objects are
PSRJ0337+1715 \citep{Ransom2014}, which is composed of two white
dwarfs and a millisecond pulsar, and WD1704+481
\citep{maxtedetal00-1}, composed of a short-period (0.15\,days)
unresolved double-degenerate and a tertiary white dwarf.

Double white dwarfs that are the inner members of hierarchical triple
systems provide another pathway towards SN\,Ia, as the tertiary can
drive Lidov-Kozai oscillations that result into their collision
\citep{Ben89, Kat12}. It is worth noting, however, that the expected
SN\,Ia rate from triples is orders of magnitude below the observed
rate \citep{Toonen2018}, which implies white dwarf collisions in
triple systems are unlikely to be the dominant channel for producing
such explosions. Lidov-Kozai driven mergers of white dwarf triples
could explain a few odd white dwarf binaries in which the more massive
component is hotter / looks younger \citep[see][]{ferrarioetal97-1,
  andrewsetal16}.

In this paper we report the discovery of J1953$-$1019, the first
resolved triple system formed by three white dwarfs.

\begin{figure}
\centerline{\includegraphics[width=0.45\textwidth]{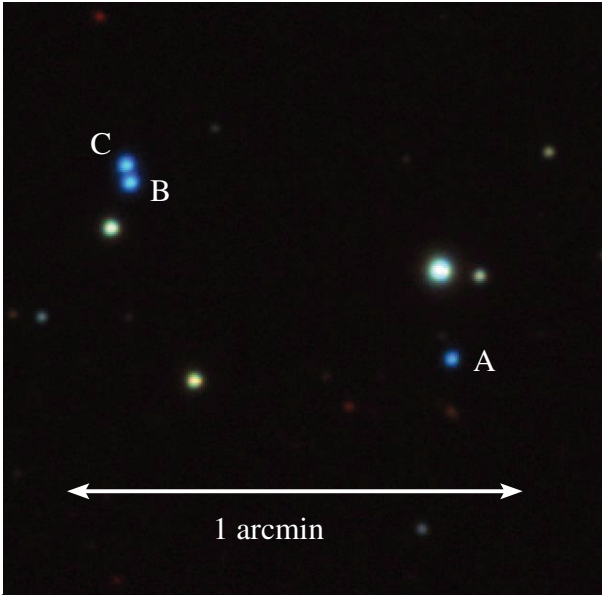}}
    \caption{\label{fig:image} PanSTARRS composite $g,i,y$ image of
      the triple white dwarf system J1953$-$1019, components A, B, and
      C (see Table\,\ref{tab:obs}) are labelled.}
\end{figure}

\section{Identification of J1953--1019}

We discovered the resolved triple white dwarf J1953$-$1019 as part of
a search for double white dwarfs in common proper motion pairs. We
performed this search by cross-matching 57\,201 objects with a
\textit{probability of being a white dwarf} greater than 0.5 from the
\citet{gentile-fusilloetal15-1} catalogue of Sloan Digital Sky Survey
(SDSS) photometric white dwarf candidates, with the \emph{Gaia} Data
Release~2 (DR2, \citealt{lindegrenetal18-1}). The \textit{Gaia}
astrometry unambiguously allows us to identify the genuine
white dwarfs among the initial list of candidates, and the cross-match
provided us with parallaxes, proper motions, and both $ugriz$ and
\emph{Gaia} photometry for a large sample of high-confidence white
dwarfs. Given that white dwarfs are nearby low-luminosity objects with
correspondingly large proper motions, care has to be taken in the
cross-match as the \textit{Gaia} and SDSS observations may have been
obtained over a decade apart. Thus, we divided our cross-matching
procedure in three separate steps. For each SDSS white dwarf candidate
we first retrieved every matching \emph{Gaia} source within a radius
of 30\arcsec\, (typically four to eight objects). We then used the
\emph{Gaia} proper motions to ``backward project'' the epoch 2015.5
\emph{Gaia} coordinates to the epoch of the SDSS observation. We
considered a true match to be the closest \emph{Gaia} source with
``backward projected'' coordinates within 2\arcsec\, of the SDSS
ones. For each white dwarf with a true match, we then selected all
\textit{Gaia} sources within 2\,\arcmin, which resulted in groups of
typically two and up to ten \emph{Gaia} targets. Within each group we
searched for white dwarfs with similar (1$\sigma$) parallaxes and
(5$\sigma$) proper motions. We allowed proper motions to agree within
a larger error to account for radial and/or orbital motions that would
have made our white dwarf common-proper-motion pair candidates in each
group seem unbound rather than co-moving objects.

This search led to the discovery of a single white dwarf triple,
J1953$-$1019. The PanSTARRS image of J1953$-$1019 clearly reveals the
three white dwarfs as blue stars (Fig.\,\ref{fig:image}). We provide
the \emph{Gaia} DR2 coordinates, parallaxes, proper motions and $G$,
$B_\mathrm{p}$, $R_\mathrm{p}$ magnitudes as well as the SDSS $g$
magnitudes of the three components in Table\,\ref{tab:obs}.

\begin{table*}
  \centering
  \setlength{\tabcolsep}{4ex}
  \caption{\textit{Gaia} coordinates, parallaxes, proper motions, $G$,
    $B_\mathrm{p}$, $R_\mathrm{p}$ and SDSS $g$ magnitudes of the
    three white dwarfs in the triple system J1953$-$1019. Masses,
    effective temperatures, surface gravities and cooling ages are
    also indicated, derived from SOAR spectroscopy. The error bars
    include external data reduction uncertainties of 1.2\% in $T_{\rm
      eff}$ and 0.038 dex in $\log g$ \citep{liebertetal05-1}. The
    most plausible range of white dwarf progenitor masses for each
    component are given in the last row.}
\begin{tabular}{cccc}
\hline
\hline
J1953$-$1019 & Component A & Component B & Component C \\
\hline
\textit{Gaia} DR2 source ID & 4190500054845023488 & 4190499986125543168 & 4190499986125543296 \\
RA (2015.5) & 19 53 33.11 & 19 53 35.99 & 19 53 36.02 \\
DEC (2015.5) & $-$10 19 55.10 & $-$10 19 31.76 & $-$10 19 29.49 \\
$\varpi$ [mas] & 7.79 $\pm$ 0.15 & 7.67 $\pm$ 0.12 & 7.64 $\pm$ 0.12 \\
Distance [pc] & 128.36 $\pm$ 2.40 & 130.37 $\pm$ 2.11 & 130.89 $\pm$ 2.01 \\
p.m. RA [mas yr$^{-1}$] & $-$10.81 $\pm$ 0.25 & $-$11.54 $\pm$ 0.18 & $-$10.94 $\pm$ 0.19 \\
p.m. DEC [mas yr$^{-1}$] & $-$16.10 $\pm$ 0.15 & $-$16.33 $\pm$ 0.11 & $-$15.66 $\pm$ 0.12 \\
$G$ [mag] & 17.28 & 16.35 & 16.44 \\
$B_\mathrm{p}$ [mag] & 17.27 & 16.05 & 16.29 \\
$R_\mathrm{p}$ [mag] & 17.30 & 16.30 & 16.44 \\
$g$ [mag] & 17.18 & 16.15 & 16.24 \\
\hline
Masses [\Msun] & 0.63 $\pm$  0.03  & 0.62  $\pm$ 0.03  &  0.60 $\pm$  0.03   \\
\Teff\, [K] & 13\,715 $\pm$ 310  &  22\,223 $\pm$ 360 &  22\,104  $\pm$ 350    \\
$\log$(g) [dex] & 8.03 $\pm$ 0.05  &  7.98 $\pm$ 0.05  &  7.95 $\pm$ 0.05   \\
Cooling age [Gyr] &  0.29 $\pm$ 0.04  & 0.042 $\pm$ 0.008  & 0.040 $\pm$ 0.008  \\
Most plausible prog. mass range [\Msun]&  1.90--2.60   &  1.80--2.55    &   1.60--2.35 \\
\hline
\end{tabular}
\label{tab:obs}
\end{table*}

\section{Spectroscopy and stellar parameters}
\label{sec:spec}

We obtained spectroscopic follow-up of all three white dwarf
components of the triple system J1953$-$1019 on the night of 2018 May
5 using the Goodman spectrograph \citep{clemens04} mounted on the
4.1-m Southern Astrophysical Research (SOAR) telescope at Cerro
Pach\'{o}n in Chile. The observations were obtained under photometric
conditions and $\simeq1.2$\,\arcsec\ seeing, using a 930 line
mm$^{-1}$ grating and a custom grating and camera angle setup that
covers the wavelength range $\simeq3600-5200$\,\AA.

Our spectroscopy of the outer tertiary (component A, see
Table\,\ref{tab:obs}) was obtained through a 3.2\,\arcsec\ slit in
order to maximize signal-to-noise (S/N) ratio. We obtained
$11\times300$\,s exposures with a resolution of roughly 3.2\,\AA\ (set
by the 1.2\,\arcsec\ seeing), achieving S/N\,=\,77 per resolution
element in the continuum at 4600\,\AA.

We subsequently used a smaller 1.0\,\arcsec\ slit to observe the inner
pair, rotating the position angle to 120\,deg so that we could obtain
a spectrum of each object individually. The narrower slit also
slightly improved the spectral resolution to $\simeq2.8$\,\AA. We
observed the northern component (C) with $4\times300$\,s exposures
(S/N\,=\,108), and the southern component (B) with $3\times300$\,s
exposures (S/N\,=\,97). The observations of the southern component
were immediately followed by an iron-argon arc lamp for wavelength
calibration; the data collected with the 3.2\,\arcsec\ slit suffer
from a large wavelength offset.

All spectra were flux calibrated using the standard star EG\,274, and
were bias- and flat-fielded using the software packages
\textsc{pamela} and \textsc{molly} \citep{marsh89}. We also applied an
absolute flux calibration to account for slit losses by normalizing
the spectra to their observed SDSS $g$-band magnitudes. The spectra of
the three white dwarfs revealed the typical broad absorption Balmer
lines of DA white dwarfs.

\begin{figure}
	\includegraphics[width=\columnwidth]{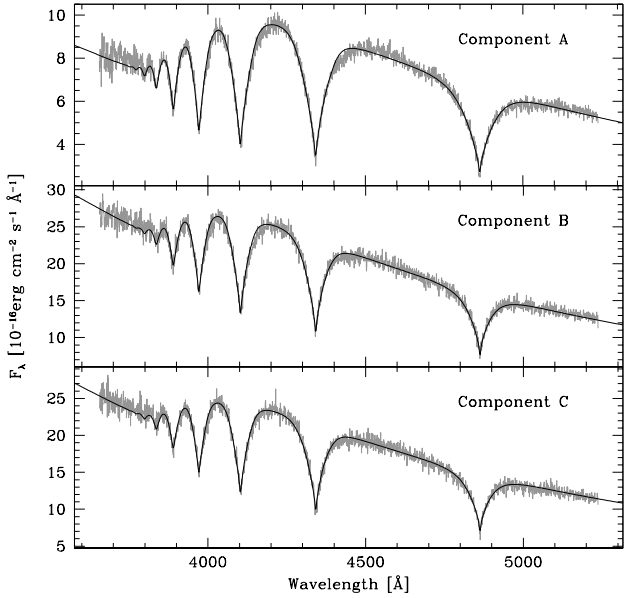}
    \caption{Best-fit models (black, see Table~\ref{tab:obs}) to the
      SOAR spectra (gray) of the three DA white dwarfs in the triple
      system J1953$-$1019.}
    \label{fig:fits}
\end{figure}

We used the fitting routine outlined in \citet{bergeronetal92-1} to
derive the surface gravities and effective temperatures of the three
white dwarfs from the SOAR spectra. To that end we used the white
dwarf model atmosphere spectra of \citet{tremblayetal11-2}, including
the Stark broadening tables of \citet{tremblay+bergeron09-1}. Given
that 1D white dwarf model spectra such as those used in this work
yield overestimated surface gravity values for white dwarfs of
effective temperatures below $\simeq13\,000$\,K
\citep{koesteretal09-1}, we compared our best-fitted effective
temperature and surface gravity values to those obtained after
applying the the 3D corrections of \citet{tremblayetal13-1} and found
no difference. Fig.\,\ref{fig:fits} shows the SOAR spectra along with
the best-fit models and the fit parameters are reported in
Table\,\ref{tab:obs}. We then derived the masses and cooling ages
(also provided in Table\,\ref{tab:obs}) interpolating the surface
gravities and effective temperatures in the cooling sequences of
\citet{fontaineetal01-1} for thick hydrogen layers and C/O-cores. We
compared the spectroscopic values thus obtained to those independently
and photometrically derived for the three white dwarfs from the
\textit{Gaia} parallaxes and photometry by \citet{Jimenez2018} and
\citet{Gentile2018} and found the values to agree within
$\simeq$2.5$\sigma$ (\Teff), $\simeq$0.3$\sigma$ ($\log$ g) and
$\simeq$0.3$\sigma$ (mass), where $\sigma$ is defined as

\begin{equation}
\sigma = \frac{|\mathrm{Spec_{value}} - \mathrm{Phot_{value}}|}{\sqrt{\mathrm{Spec_{error}}^2+\mathrm{Phot_{error}}^2}},
\notag
\end{equation}
\noindent and Spec$_\mathrm{value}$, Spec$_\mathrm{error}$,
Phot$_\mathrm{value}$, Phot$_\mathrm{err}$ are the spectroscopic and
photometric stellar parameter values and their errors.

\section{Progenitor masses}
\label{sec:mass}

It has been shown that wide white dwarf binaries can be used to
constrain the initial-to-final mass relation, since the difference
between white dwarf cooling ages gives an indication of the difference
between progenitor lifetimes, assuming the binary components are
coeval \citep{finley+koester97-1, girvenetal10-1, Baxter2014,
  andrewsetal15-1}.  In this section we test if the parameters we
derive for J1953$-$1019 are consistent with coeval evolution, when
adopting the initial-to-final mass relations of
\citet{ferrarioetal05-1} (their non-linear relation),
\citet{Catalan2008a} (their two-piece relation),
\citet{kaliraietal08-1}, \citet{williamsetal09-1},
\citet{salarisetal09-1} (their linear relation),
\citet{renedoetal10-1}, \citet{gesickietal14-1} (their
  equation 15) and \citet{cummingsetal18-1} (their MIST-based
  relation).

From the above relations we first obtained the progenitor masses of
the three components, where the final masses are the white dwarf
masses reported in Table\,\ref{tab:obs}. In this exercise we took into
account the white dwarf mass uncertainties, hence we obtained a
possible range of progenitor masses from each relation. The results
are given in Table\,\ref{tab:masses}. For each range of progenitor
masses we then derived the corresponding main sequence lifetimes using
a main sequence lifetime-mass relation. These values are also provided
in Table\,\ref{tab:masses}. We obtained a main sequence lifetime-mass
relation by employing the main sequence stellar evolutionary tracks of
\citet{Choietal16}, where we adopted solar metallicities and
  rotation at 0.4 of critical velocity \citep[see
  also][]{dotter16}. Specifically, we computed tracks for masses
ranging from 1\,\Msun\, to 10\,\Msun\ in steps of 0.1\,\Msun, which
translated into main sequence lifetimes from $\simeq0.1$\,Myr to
$\simeq10$\,Gyr. From each track we extracted the time (i.e. age or
main sequence lifetime) and the mass of the star corresponding to the
transition into a white dwarf, therefore having a function relating
the main sequence lifetime and the main sequence mass.

\begin{figure}
	\includegraphics[width=\columnwidth]{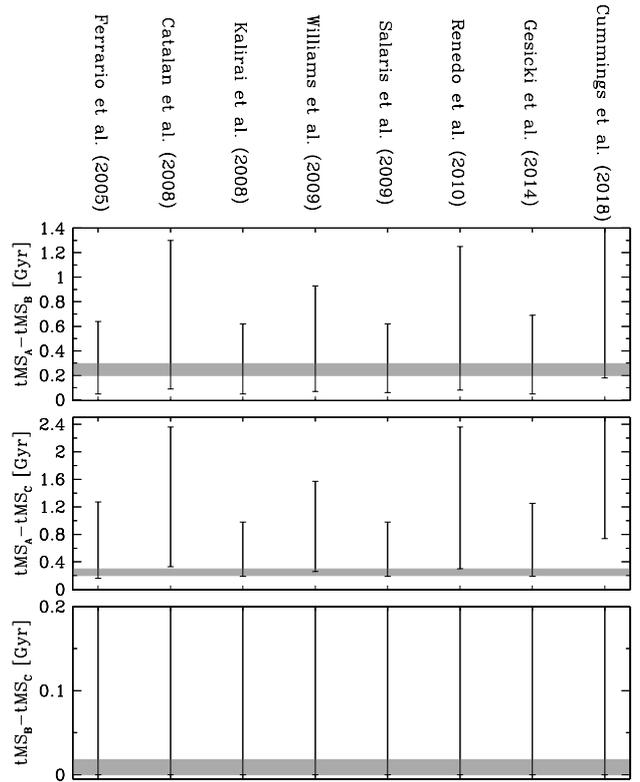}
    \caption{The possible range of values (black vertical lines) we
      derive for the difference between the main sequence progenitor
      lifetimes (obtained assuming eight different initial-to-final
      mass relations) as compared to the difference in cooling times
      implied by the atmospheric parameters that we derived from the
      SOAR spectroscopy, indicated by shaded grey regions. The top,
      middle and bottom panels illustrate the time differences between
      components A--B ($t_\mathrm{MS_A}-t_\mathrm{MS_B}$), A--C
      ($t_\mathrm{MS_A}-t_\mathrm{MS_C}$) and B--C
      ($t_\mathrm{MS_B}-t_\mathrm{MS_C}$), respectively. Note that the
      difference in cooling ages is very small between components B
      and C, we hence zoom-in to this region in the bottom panel, and
      consequently the ranges derived from the initial-to-final mass
      relations extend off the scale.}
    \label{fig:image3}
\end{figure}

From all the possible values obtained by computing
$t_\mathrm{MS_A}-t_\mathrm{MS_B}$, $t_\mathrm{MS_A}-t_\mathrm{MS_C}$
and $t_\mathrm{MS_B}-t_\mathrm{MS_C}$ (where $t_\mathrm{MS_A}$,
$t_\mathrm{MS_B}$ and $t_\mathrm{MS_C}$ are the main sequence
lifetimes reported in Table\,\ref{tab:masses}), we kept the maximum
and the minimum in each case, which defined the maximum and minimum
differences between the progenitor lifetimes of the three
components. These differences are compared to the difference in white
dwarf cooling times in Fig.~\ref{fig:image3} for each of our adopted
initial-to-final mass relations.

\begin{table*}
  \centering
  \setlength{\tabcolsep}{3ex}
  \caption{The possible range of white dwarf progenitor masses derived
    adopting eight initial-to-final mass relations and taking into
    account the white dwarf masses (and their uncertainties) reported
    in Table\,\ref{tab:obs}. The ranges of main sequence lifetimes are
    also indicated.}
\begin{tabular}{ccccccc}
\hline
\hline
IFMR                     & Comp. A     & Comp. A   & Comp. B    & Comp. B   & Comp. C    & Comp. C \\
                         & Prog. mass  & MS lifet. & Prog. mass & MS lifet. & Prog. mass & MS lifet. \\
                         &   [\Msun]   & [Gyr]     & [\Msun]    & [Gyr]     & [\Msun]    & [Gyr] \\
\hline
\citet{ferrarioetal05-1} & 2.08--2.61 & 0.60--1.06 & 1.97--2.53 & 0.65--1.24 & 1.72--2.36 & 0.76--1.87  \\
\citet{Catalan2008a}     & 1.78--2.40 & 0.72--1.68 & 1.68--2.30 & 0.81--2.02 & 1.47--2.09 & 1.05--3.08  \\
\citet{kaliraietal08-1}  & 2.02--2.49 & 0.67--1.15 & 1.95--2.41 & 0.72--1.29 & 1.79--2.26 & 0.86--1.65  \\
\citet{williamsetal09-1} & 1.89--2.44 & 0.70--1.40 & 1.80--2.35 & 0.77--1.63 & 1.61--2.16 & 0.96--2.27  \\
\citet{salarisetal09-1}  & 2.01--2.45 & 0.69--1.18 & 1.93--2.38 & 0.75--1.31 & 1.78--2.23 & 0.88--1.67  \\
\citet{renedoetal10-1}   & 1.80--2.39 & 0.73--1.64 & 1.69--2.30 & 0.81--1.98 & 1.47--2.10 & 1.03--3.09  \\
\citet{gesickietal14-1}  & 2.01--2.53 & 0.65--1.17 & 1.92--2.45 & 0.70--1.34 & 1.71--2.27 & 0.84--1.90  \\
\citet{cummingsetal18-1} & 1.39--2.14 & 0.99--3.71 & 1.26--2.01 & 1.17--5.07 & 1.01--1.76 & 1.73--10.7  \\
\hline
\end{tabular}
\label{tab:masses}
\end{table*}

Inspection of the top panel of Fig.~\ref{fig:image3} reveals that the
difference in cooling ages and main sequence lifetimes between
components A and B agree when considering all initial-to-final mass
relations. The same is true in the middle panel of the same figure
when comparing the time differences between the A and C components,
except for the \citet{Catalan2008a}, \citet{renedoetal10-1} and
\citet{cummingsetal18-1} relations. The bottom panel of
Fig.~\ref{fig:image3} indicates that, for components B and C, all the
considered relations provide differences in main sequence lifetimes
which are also in agreement with those obtained from the cooling
times.

The above exercise shows that the masses and cooling ages of the three
white dwarfs in J1953$-$1019 are consistent with a coeval evolution
when we consider the initial-to-final mass relations of
\citet{ferrarioetal05-1}, \citet{kaliraietal08-1},
\citet{williamsetal09-1}, \citet{salarisetal09-1}, and
\citet{gesickietal14-1}. The most plausible white dwarf progenitor
masses should be within the ranges given in Table\,\ref{tab:masses}
for these relations, i.e. 1.90--2.61\,\Msun\, for component A,
1.80--2.53\,\Msun\, for component B and 1.61--2.36\,\Msun\, for
component C. These most plausible mass ranges for the individual
initial-to-final mass relations are provided in Table\,\ref{tab:obs}.

It is important to emphasise that our aim here is to check whether or
not the stellar parameters we derived for the three white dwarfs in
the triple system J1953$-$1019 are consistent with coeval evolution
and to estimate their progenitor masses. Our results should not be
used to discriminate or favour a specific initial-to-final mass
relation.

We also need to bear in mind that we calculated the main sequence
progenitor lifetimes using the evolutionary models of
\citet{Choietal16} and that the initial-to-final mass relations
assumed in this work were derived adopting different evolutionary
models than ours, namely those by \citet{hurleyetal00-1,
  girardietal02-1, pietrinfernietal04-1, marigo+girardi07-1,
  bressanetal12-1}. In order to check the impact of the choice of
evolutionary model in our results we re-calculated the differences
between the progenitor lifetimes of the three white dwarfs employing
these evolutionary tracks. We found that, even though the calculated
progenitor lifetimes varied slightly from one model to another, the
time differences were in agreement with those obtained from the
cooling times for the \citet{ferrarioetal05-1},
\citet{kaliraietal08-1}, \citet{williamsetal09-1},
\citet{salarisetal09-1}, and \citet{gesickietal14-1} initial-to-final
mass relations. We hence conclude the three white dwarfs in
J1953$-$1019 are consistent with being co-eval independently of the
evolutionary models used.

\section{Orbital Parameters}
\label{sec:orbit}

The triple system J1953$-$1019 consists of an inner white dwarf binary
and a wider white dwarf companion (Fig.\,\ref{fig:image}). We derived
an angular separation of 2\,332.69\,mas (the uncertainty is of the
order of $10^{-6}$ mas) between the inner two white dwarfs from their
\emph{Gaia} coordinates (Table\,\ref{tab:obs}). To obtain the angular
separation between the inner white dwarf binary and the wider
companion we calculated the distance and coordinates of the centre of
mass (CoM) of the inner binary ($d$\,=\,130.62\,pc; RA\,=\,298.40003
deg, Dec\,=\,$-$10.32518 deg), which combined with the \emph{Gaia}
coordinates of the wider component (component A; Table\,\ref{tab:obs})
resulted in an angular separation of 49\,222.83\,mas (the uncertainty
is of the order of $10^{-4}$ mas). Taking 130\,pc as the distance to
the triple white dwarf (obtained simply as the weighted average of the
three distances to each white dwarf component calculated from
inverting the measured parallaxes), we obtained a projected separation
of the inner binary of $303.25\pm0.01$\,au and of
$6\,398.97\pm0.09$\,au for the inner binary and the outer companion.

Knowing the angular separations as well as the distances to each
individual star and the CoM, one could attempt deriving both the true
orbital separation between the inner binary and between the CoM of the
inner binary and the outer companion. However, we do not proceed with
this exercise for the following reason: while the \textit{Gaia}
positions of the two stars are extremely accurately measured by {\em
  Gaia}, the parallax uncertainties, though remarkably small
($\simeq2$\%), are of the order of the orbital separations of the
stars in J1953$-$1019, and do hence not constrain the orbital
separations. For this reason, in what follows we only make use of the
calculated projected separations.

\section{Evolution}
Here we assess the formation and evolution of J1953--1019. We adopt
the masses from Sect.\,\ref{sec:mass} and projected separations from
Sect.\,\ref{sec:orbit}, however the system is under-constrained; that
is, the eccentricities and mutual inclinations are unknown. Therefore,
we take seven configurations that span the possible parameter space in
which J1953--1019 is currently dynamically stable
(Table\,\ref{tbl:conf}). We take the projected separations as the
current distance between the inner two stars, and that of the outer
star to the CoM of the inner binary. For a given eccentricity of an
orbit, we derive a maximum and minimum orbital separation assuming the
system is currently at pericentre or apocentre. To test the dynamical
stability of this three-body system, we apply the criterion of
\citet{Mar99}:

\begin{eqnarray}
\begin{array}{l c l}
\dfrac{a_{\rm out}}{a_{\rm in}}|_{\rm crit} &= &\dfrac{2.8}{1-e_{\rm out}} (1-\dfrac{0.3i}{\pi}) \cdot \\
&&\\
&&\left( \dfrac{(1.0+q_{\rm out})\cdot(1+e_{\rm out})}{\sqrt{1-e_{\rm out}}} \right)^{2/5}, \\
 \end{array} 
\label{eq:stab_crit}
\end{eqnarray}
where $q_{\rm out}\equiv \dfrac{m_{\rm A}}{m_{\rm B}+m_{\rm C}}$ with
$m_{\rm A}, m_{\rm B}, m_{\rm C}$ the masses of components A, B and C
respectively. Furthermore $a_{\rm in}$ and $a_{\rm out}$ are the
semi-major axis of the inner and outer orbit, and $e_{\rm in}$ and
$e_{\rm out}$ the corresponding eccentricities, and $i$ is the mutual
inclination. Systems are stable if $\dfrac{a_{\rm out}}{a_{\rm in}} >
\dfrac{a_{\rm out}}{a_{\rm in}}|_{\rm crit}$.  For J1953$-$1019 the
condition for stability is met for circular orbits, but breaks down
for very eccentric orbits, for example $[e_{\rm in} =0$, $e_{\rm out}
  >0.63$, $i =0^{\circ}]$ or $[e_{\rm in} >0.85$, $e_{\rm out} =0$, $i
  =0^{\circ}]$. The last configuration listed in Table\,\ref{tbl:conf}
is dynamically stable only for retrograde orbits with mutual
inclinations above 128$^{\circ}$. If the system is dynamically
unstable, it is likely to dissolve on a timescale of a few times the
inner orbit. For this reason we only consider dynamically stable
systems with $e \leq 0.5$.

Next we investigate whether the presence of the tertiary star affects
the orbit of the inner binary. The lowest-order manifestation of the
interaction are the Lidov-Kozai cycles \citep{Koz62, Lid62}, in which
the eccentricity of the inner orbit and the mutual inclination varies
periodically.  The timescale of these cycles is \citep{Kin99}:
\begin{equation}
t_{\rm kozai} \approx \dfrac{P_{\rm out}^2}{P_{\rm in}} \dfrac{m_{\rm A}+m_{\rm B}+m_{\rm C}}{m_{\rm A}} \left(1-e_{\rm out}^2\right)^{3/2},
\label{eq:t_kozai}
\end{equation}
where $P_{\rm in}$ and $P_{\rm out}$ are the periods of the inner and
outer orbit, respectively. If the inner and outer orbits are circular,
the Lidov-Kozai timescale is $\simeq 90$\,Myr. With eccentric orbits
($e \leq 0.5$, Table\,\ref{tbl:conf}), the range of Lidov-Kozai
timescales broadens ranging from a few to several hundred Myr. During
the pericenter passage of the inner binary during the
high-eccentricity phase of the Lidov-Kozai cycle, the inner binary can
be driven towards mass transfer or a collision. This collision may
result in sub-Chandrasekhar (since the total mass of the inner binary
is slightly larger than 1.2M$_\mathrm{\odot}$) SN\,Ia explosion
\citep[e.g.][]{shenetal17-1}.

For the regular Lidov-Kozai cycles, the maximum amplitude in
eccentricity $e_{\rm max}$ is strongly dependent on the unknown mutual
inclination. For example, in the test-particle approximation, $e_{\rm
  max} = \sqrt{1-\frac{5}{3} \mathrm{cos ^2(i_{\rm i})}}$
\citep[e.g.][]{Nao13} and Lidov-Kozai cycles can only take place when
the mutual inclination is between 39--141$^{\circ}$.

The importance of higher-order terms of the three-body approximation,
such as eccentric Lidov-Kozai cycles, can be quantified with the
octupole parameter \citep{Nao16}
\begin{equation}
\epsilon_{\rm oct} = \dfrac{m_{\rm B}-m_{\rm C}}{m_{\rm B}+m_{\rm C}} \dfrac{a_{\rm in}}{a_{\rm out}} \dfrac{e_{\rm out}}{1-e_{\rm out}^2}
\notag
\end{equation}
As $|\epsilon_{\rm oct}| \ll 0.01$, the eccentric Lidov-Kozai
mechanism is not likely important for the majority of possible
configurations of J1953$-$1019.

The chance for a collision to occur increases if the system is in the
quasi-secular regime of dynamical evolution. In this state significant
pericentre changes occur on a single orbit timescale, such that the
pericentre approach can be arbitrarily close leading to physical
collisions \citep{Ant12, Kat12}. A triple is in the quasi-secular
regime if \citep{Ant14}:
\begin{equation}
\sqrt{1-e_{\rm in}} \gtrsim 5\pi \frac{m_{\rm A}}{m_{\rm B}+m_{\rm C}} \left ( \frac{a_{\rm in}}{a_{\rm out}(1-e_{\rm out})} \right)^3 
\notag
\end{equation}
For this to occur, the inner eccentricity needs to reach extremely
high values during the three-body interactions; for most orbital
configurations from Table\,\ref{tbl:conf}, $e_{\rm max} \gtrsim 0.97$,
and $e_{\rm max} \gtrsim 0.7$ for the last system. Equivalently, if
the regular Lidov-Kozai cycles are to drive the system to high
eccentricities and subsequent collisions in the inner orbit, the
mutual inclinations should lie between $80-110^{\circ}$ and down to
$56-124^{\circ}$ for the last system.

In conclusion, there are rare, favourable conditions of the current
eccentricities and mutual inclinations that make it possible for the
inner binary to experience a collision due to the dynamical
interaction with the tertiary white dwarf, making J1953$-$1019 a
possible (sub-Chandrasekhar) SN\,Ia progenitor. Failing to ignite,
this system would turn into a wide double white dwarf binary with
inconsistent masses and cooling ages, akin to RE\,J0317--853
\citep{kulebietal10-1} and HS\,2220+2146
\citep{andrewsetal16}. However, for the majority of the parameter
space of possible orbital configurations, the inner and outer orbit of
J1953$-$1019 are effectively decoupled.

Regarding the formation of the white dwarfs, it is likely that mass
transfer has not taken place in this system and that the stars evolved
into white dwarfs effectively as single stars. The reason is that if
mass transfer would have taken place, the orbit would likely have been
circularized due to tides, and the inner orbital separation would be
equal to the observed projected separation of 303.25\,au; this is much
wider than expected for post-mass transfer systems \citep[see
  e.g.][]{Too14}. We note that the three-body effects that potentially
can lead to the collision of the white dwarfs should not be very
efficient before the formation of the white dwarfs in order to avoid
mass transfer. The current configuration of J1953$-$1019 also implies
that the system has survived flybys that could have potentially led to
unbinding the outer white dwarf \citep{Antognini2016}.

\begin{table}
\centering
\caption{Seven possible configurations that span the possible
  parameter space in which J1953--1019 is currently dynamically stable
  and their related timescales.}
\begin{tabular}{llll|ll}
\hline \hline
$e_{\rm in}$ &Current &$e_{\rm out}$&Current &$t_{\rm kozai}$ &$\epsilon_{\rm oct}$\\
           & orbital &         &orbital  & (Myr) & \\
           & phase &         & phase &  & \\
 \hline  
 0 & -            & 0 & -               & 87   & -\\
 0 & -            & 0.5 & Apocentre    & 26    & 0.0008\\
 0 & -            & 0.5 & Pericentre   & 697   & 0.0003\\
 0.5& Apocentre   & 0 & -              & 103   & -\\
 0.5& Pericentre  & 0 & -              & 20    & - \\
 0.5 & Apocentre & 0.5 & Pericentre  & 831   & 0.0002\\
 0.5 & Pericentre  & 0.5 & Apocentre & 6     & 0.0016\\
\hline 
\end{tabular}
\label{tbl:conf}
\end{table}

\section{Conclusions}

We have presented the discovery of the first spatially resolved triple
system composed of three white dwarfs. Follow-up optical spectroscopy
has allowed us to derive DA spectral types of the white dwarfs, as
well as to measure their effective temperatures, surface gravities,
masses and cooling ages. We have adopted eight previously published
initial-to-final mass relations and derived the range of possible
progenitor masses for the three components and their expected main
sequence lifetimes. The differences between these main sequence
lifetimes and the cooling ages agree for most of the adopted
initial-to-final mass relations, hence we conclude the three white
dwarfs are consistent with coeval evolution. Based on the $Gaia$~DR2
astrometry we have calculated the projected separations between the
inner binary and between the centre of mass of the inner binary and
the outer companion. Using these values, together with the masses of
the white dwarfs, we have investigated a wide range of future
evolutionary paths of the triple system. Although we cannot completely
rule out the possibility for a collision between the inner binary, we
consider such an event to be unlikely. A hypothetical collision could
lead to a sub-Chandrasekhar Type Ia supernova explosion.

\section*{Acknowledgements}
This work was supported by the MINECO Ram\'on y Cajal programme
RYC-2016-20254, by the MINECO grant AYA\-2017-86274-P, by the AGAUR
(SGR-661/2017), by the Netherlands Research Council NWO (grant VENI
[nr. 639.041.645]), by the European Research Council under the
European Union's Seventh Framework Programme (FP/2007--2013) / ERC
Grant Agreement n. 320964 (WDTracer), by Horizon 2020 research and
innovation programme n. 677706 (WD3D) and by NASA through Hubble
Fellowship grant \#HST-HF2-51357.001-A, awarded by the Space Telescope
Science Institute, which is operated by the Association of
Universities for Research in Astronomy, Incorporated, under NASA
contract NAS5-26555. Based on observations obtained at the Southern
Astrophysical Research (SOAR) telescope, which is a joint project of
the Minist\'{e}rio da Ci\^{e}ncia, Tecnologia, e Inova\c{c}\~{a}o da
Rep\'{u}blica Federativa do Brasil, the U.S. National Optical
Astronomy Observatory, the University of North Carolina at Chapel
Hill, and Michigan State University.





\bsp	
\label{lastpage}
\end{document}